\documentclass[twocolumn,10pt,aps,prd]{revtex4-2}
\usepackage{hyperref,graphicx,amsfonts,amsmath,amssymb}
\usepackage[latin1]{inputenc}
\usepackage{multirow}
\usepackage{subcaption}
\usepackage{diagbox}
\newcommand{\tn}{\textnormal}
\begin{document}
\title{Heavy neutral 2HDM Higgs Boson Pair Production at CLIC Energies }

\author{Majid Hashemi}
\email{majid.hashemi@cern.ch}
\author{Marieh Molanaei}
\email{mmovlanaei@gmail.com}
\affiliation{Physics Department, College of Sciences, Shiraz University, Shiraz, 71946-84795, Iran}

\begin{abstract}
In this work, the neutral Higgs boson pair production is analyzed at Compact Linear Collider (CLIC) to be operating at center of mass energies $ \sqrt{s}=1400 $ GeV (stage 2) and $ \sqrt{s}=3000 $ GeV (stage 3). The Higgs bosons to be searched for are neutral CP-even ($H$) and CP-odd ($A$) within the framework of two Higgs doublet model (2HDM) in the mass range $300<m_{H/A}<1000$ GeV. All types of the CP-conserving model are studied and the signal observability is evaluated taking into account the main SM background processes like $ZZ$, $t\bar{t}$ and the SM-like Higgs boson associated production ($hZ$). Results are presented for a set of model parameters and Higgs boson masses in terms of signal distributions over the background as well as the integrated luminosity needed for $5\sigma$ discovery. It is shown that the heavy mass region is well observable at CLIC in types 3 (flipped) and 4 (lepton-specific) in the regions not excluded by LHC so far, while in type 1 the signal observation is challenging due to the large jet multiplicity in the $t\bar{t}$ final state.
\end{abstract}
\maketitle
\section*{Introduction}
Over the past few decades, Standard Model of particle physics (SM) has been the most reasonable theory of sub-atomic interactions between elementary particles. The underlying framework which is based on quantum field theory (QFT), describes the strong, electromagnetic and weak interactions. 

One of the biggest achievements of the Large Hadron Collider (LHC) is the observation of a new boson which is considered to be the candidate for the missing element of SM, i.e., the Higgs boson introduced through the Higgs mechanism \cite{Higgs1,Higgs2,Higgs3,Englert1,Kibble1,Kibble2}. The announcement by the two collaborations CMS and ATLAS \cite{HiggsATLAS,HiggsCMS} at the Large Hadron Collider (LHC) experiment led to the Physics Nobel Prize in 2013. 

The properties of the observed particle has been verified in various analyses by CMS (\cite{LHC5,LHC6,LHC7,LHC8}) and ATLAS (\cite{LHC1,LHC2,LHC3,LHC4}). Despite the overall agreement of the measurements with SM predictions, there are still several open problems which motivate theories beyond SM (BSM) including matter-antimatter asymmetry \cite{M_AM_ASYMM1}, the origin of dark matter which is not contained in SM \cite{ColliderDM1, ColliderDM2}, the senstivity of the muon anomalous magnetic moment to new physics and its current 3 to 4 standard deviation from SM prediction \cite{muon}, the hierarchy problem, i.e., the large difference between the Higgs boson mass (125 GeV) and the Planck mass ($10^{19}$ GeV), etc. 

The stability of the observed hierarchy with respect to the quantum corrections to the Higgs boson mass is a motivation for below TeV-scale supersymmetry \cite{Martin, Aitchison}. The theory of supersymmetry provides a natural solution to the hierarchy problem by introducing super-particles which contribute to counter-terms in the perturbative series of corrections to the Higgs boson mass. In its minimal form, MSSM needs an extended Higgs sector beyond SM structure \cite{Aitchison}. It turns out that for a supersymmetric theory to work, at least two Higgs doublets are needed to give masses to the particles and their super-partners \cite{HTPH, habersusy}. This is not, of course, the only reason for extended Higgs sectors but is considered as one of the main motivations for the two Higgs doublet model (2HDM) \cite{MSSM2,MSSM3} which was first introduced as a model for CP violation and natural flavor conservation \cite{2hdm1, 2hdm2, 2hdm3}. Due to the more degrees of freedom in the model, 2HDM describes the flavor physics observables \cite{Misiak,FMahmoudi} and electroweak precision measurements \cite{Wmass} better than SM.

The building blocks of the Higgs sector of the two Higgs doublet model consist of two complex scalar doublets $\Phi_1 $ and $\Phi_2 $ with their corresponding vacuum expectation values, $v_1$ and $v_2$ under the requirement of $\sqrt{v_1^2+v_2^2}=246$ GeV \cite{2hdm_HiggsSector1,2hdm_HiggsSector2}. This is a natural expansion of the SM Higgs sector to two doublets. One of direct consequences of having two complex doublets is the multiplicity of the Higgs bosons: two CP-even scalars $h,H$, a CP-odd pseudoscalar $A$ and the charged Higgs bosons $H^{\pm}$. 

At the alignment limit \cite{align1,align2,align3}, the 2HDM partially aligns with SM in the sense that one of the Higgs bosons acquires the same properties as the SM Higgs boson. This alignment is to confirm the observation of the new boson at LHC which is usually assumed to be the lightest 2HDM scalar ($h_{SM}$), although there is the so called twisted scenario where the aligned boson is not the lightest \cite{cms1}. 

In a CP-conserving scenario, there are four types of the model characterized by the Higgs boson couplings with up/down-type quarks and leptons which are functions of the ratio of the vacuum expectation values denoted by $\tan\beta$. Therefore $\tan\beta$ plays the key role in 2HDM phenomenology \cite{tanbsignificance}.

The SM Higgs boson studies are going on at LHC including pair production ($hh$) in different final states by CMS \cite{CMSHH1,CMSHH2,CMSHH3,CMSHH4,CMSHH5} and ATLAS \cite{ATLASHH1,ATLASHH2,ATLASHH3,ATLASHH4}. However 2HDM Higgs bosons are searched for in single production mode by CMS \cite{cms1,cms2,cms3,cms4} and ATLAS \cite{atlas1,atlas2,atlas3}. The above scenarios are based on $A\to Zh$, $A \to ZH$, $H \to ZA$ or $H\to hh$ with the mother particle produced in proton-proton collisions. The Higgs conversion modes studied by both CMS and ATLAS are limited to the kinematic requirement $|m_A-m_{h/H}|>m_Z$. Therefore, there are open regions in the 2HDM parameter space specially along the equal masses line $m_A=m_H$ favored by $\Delta\rho$ requirements \cite{drho1,drho2,drho3,drho4}.    

While LHC analyses continue to search for extra Higgs bosons, other experimental scenarios are also under consideration including LHC luminosity upgrade \cite{hllhc1,hllhc2} and future lepton colliders, CLIC \cite{clichiggs1,clichiggs2}, ILC \cite{ilchiggs}, FCC \cite{fcchiggs} and CEPC \cite{cepchiggs}.
 
The lepton colliders are of interest due to \\

-providing cleaner collision event environment and lower QCD background due to the less hadron activity compared to hadron colliders, \\

-better knowledge of the effective center of mass energy and the beam spectrum compared to hadron collisions where parton distribution functions play an important role and \\

-kinematic constraints which make event reconstruction easier. The latter is exploited in the current analysis and will be described in more details.

In recent works, several search scenarios were introduced within type 1 \cite{HG_2,HG_3,HG_4,HE}, type 3 \cite{HG_1,HN} and type 4 \cite{MH_1,MH_2,HG_5}. These analyses are dedicated to lepton colliders but do not cover the heavy neutral Higgs boson masses beyond 500 GeV. In the current analysis, we explore heavy neutral Higgs bosons with masses up to 1 TeV at CLIC center of mass energies of 1.4 and 3 TeV known as stages 2 and 3 respectively. Details of the analysis will be presented after a theoretical introduction. 

	
\section{Theoretical Framework}
Consider the scalar Higgs sector of the SM Lagrangian as described in the Higgs mechanism:
\begin{eqnarray}\label{potential-SM}
	\mathcal{L}_{\Phi}^{\textnormal{SM}}&=&(D^{\mu}\Phi)^{\dagger}(D_{\mu}\Phi)-\mathcal{V}^{\textnormal{SM}}(\Phi)
\end{eqnarray}
where $D_\mu$ is the covariant derivative and $\Phi$ is the complex scalar doublet in the spinor representation of $SU(2)_L$:
\begin{eqnarray}\label{Standard Model}
\Phi=\binom{\phi^{+}}{\phi^{0}}
\end{eqnarray}
The renormalizable Higgs potential using one doublet takes the form:
\begin{eqnarray}\label{potential-SM}
\mathcal{V}^{\textnormal{SM}}&=&-\mu^{2}\Phi^{\dagger}\Phi+\lambda (\Phi^{\dagger}\Phi)^2 
\end{eqnarray}	
with $\mu$ and $\lambda$ as the free parameters of the model describing the mass term and quartic self interactions among scalars. In this form, $\lambda$ is required to be positive for the vacuum stability and a non-zero vev is obtained provided that $\mu^2>0$. 

Extensions of SM have been extensively studied in the literature, including real scalar extension \cite{RealxSM1,RealxSM2,RealxSM3,RealxSM4,RealxSM5,RealxSM6,RealxSM7}, complex scalar extension \cite{Complex1,Complex2,Complex3,Complex4,Complex5}, inert doublet model \cite{IDM1,IDM2,IDM3,IDM4,IDM5} and inert doublet plus a complex scalar \cite{IDMCS}. These scenarios try to address the two problems of the nature of the dark matter and the origin of the baryon asymmetry. The two Higgs doublet model is also one of the simplest SM extensions with an additional complex scalar doublet:
\begin{eqnarray}\label{Standard Model}
\Phi_{i}=\binom{\phi_{i}^{+}}{(v_{i}+\rho_i+i\eta_i)/\sqrt{2}},~ i=1, 2.  
\end{eqnarray}
and the Lagrangian:
\begin{eqnarray}\label{lagrangian-2HDM}
\mathcal L_{\Phi}^{\textnormal{2HDM}}&=& \sum_{i=1,2}(D_{\mu}\Phi_{i})^{\dagger}(D^{\mu}\Phi_{i})-\mathcal V^{\textnormal{2HDM}}
\end{eqnarray}
with a symmetric potentail including mass terms and Higgs-self interactions as follows: 
\begin{align}
	\mathcal{V}^{\textnormal{2HDM}} \nonumber &= m_{11}^2\Phi_1^\dagger\Phi_1+m_{22}^2\Phi_2^\dagger\Phi_2-m_{12}^2\left(\Phi_1^\dagger\Phi_2+\Phi_2^\dagger\Phi_1\right)\\
	\nonumber &+\frac{1}{2}\lambda_1\left(\Phi_1^\dagger\Phi_1\right)^2+\frac{1}{2}\lambda_2\left(\Phi_2^\dagger\Phi_2\right)^2\\ \nonumber
	\nonumber &+\lambda_3\left(\Phi_1^\dagger\Phi_1\right)\left(\Phi_2^\dagger\Phi_2\right)+\lambda_4\left(\Phi_1^\dagger\Phi_2\right)\left(\Phi_2^\dagger\Phi_1\right)\\ \nonumber
	\nonumber &+\frac{1}{2}\lambda_5\left[\left(\Phi_1^\dagger\Phi_2\right)^2+\left(\Phi_2^\dagger\Phi_1\right)^2\right].\\ 
\end{align}
The neutral Higgs bosons are obtained through a rotation in the $\rho_i$ and $\eta_i$ spaces with mixing angles $\alpha/\beta$ for the CP-even/odd Higgs bosons:
\begin{align}
h\nonumber&=-\rho_1\sin\alpha+\rho_2\cos\alpha \\ \nonumber
H\nonumber&=\rho_1\cos\alpha+\rho_2\sin\alpha \\ \nonumber
A\nonumber&=\eta_1\sin\beta+\eta_2\cos\beta  \nonumber
\label{mixing}
\end{align}
while the charged Higgs is related to the charged fileds $\phi^{\pm}$:
\begin{equation}
	H^{\pm}=\phi^{\pm}_1\sin\beta+\phi^{\pm}_2\cos\beta.
\end{equation}
The mixing angle $\beta$ is also related to the ratio of vev's through:
\begin{eqnarray}\label{parameter1}
\tan \beta&=&\frac{v_{2}}{v_{1}}
\end{eqnarray}
The Yukawa Lagrangian for the Higgs-fermion interaction with fermions follows the usual form for CP-even Higgs bosons ($h,~H$) while for CP-odd Higgs boson $A$ takes additional $i\gamma^{5}$ in the vertex:
\begin{eqnarray}\label{lagrangian2HDM}
\mathcal L_{\Phi}&=& \frac{m_{f}}{v}\sum_{f=u,d,l}\bar{f}f(h+\rho_H^f H-i\rho_A^f\gamma^{5}A). 
\end{eqnarray}
This is the form of Higgs-fermion interactions at the alignment limit ($\beta-\alpha=\pi/2$) where as is seen the light Higgs coupling with fermions is equivalent with SM predictions. The heavy Higgs boson coupling with fermions, however, deviates from SM form of $m_f/v$ by the $\rho^f_{H/A}$ values. The $\rho^f_H$ takes the following forms for different types of the model as presented in Tab. \ref{hf}. The $\rho^f_A$ equals $\rho^f_H$ but takes an extra minus sign for $f=u$.
\begin{table}[h]
		\centering
		\begin{tabular}{|l|cccc|}
			\hline
			\multirow{2}{*}{
	\diagbox{Coupling}{Type}} & 1 & 2 & 3 & 4\\
     & & & & \\
     \hline
		$~~~~~\rho^u$ & $\cot\beta$ & $\cot\beta$ & $\cot\beta$ & $\cot\beta$ \\
		\hline
		$~~~~~\rho^d$ & $\cot\beta$ & $-\tan\beta$ & $-\tan\beta$ & $\cot\beta$ \\
        \hline
		$~~~~~\rho^d$ & $\cot\beta$ & $-\tan\beta$ & $\cot\beta$ & $-\tan\beta$ \\
        \hline
		\end{tabular}
		\caption{Higgs-fermion couplings in different types of 2HDM at the alignment limit.}
		\label{hf}
	\end{table}
The Higgs-gauge couplings scaled to their corresponding SM values are  
\begin{eqnarray}\label{coupling1}
\frac{g_{hVV}^{\tn{2HDM}}}{g_{hVV}^{\tn{SM}}}&=& \sin(\beta-\alpha) \nonumber\\
\frac{g_{HVV}^{\tn{2HDM}}}{g_{hVV}^{\tn{SM}}}&=& \cos(\beta-\alpha).
\end{eqnarray}
The alignment limit requires $\sin(\beta-\alpha)=1$, thus leading to gaugophobic heavy Higgs bosons while the light 2HDM Higgs boson fully coincides with SM prediction in terms of the couplings with fermions and gauge bosons. 

\section{Cross section and decay rates}
\begin{figure}[h!]
	\includegraphics[width=0.8\linewidth]{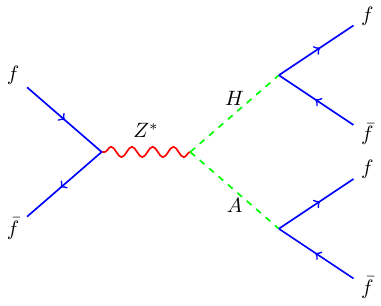}
	\caption{Feynman diagram of the Higgs boson pair production in the four fermion final state. The fermion $f$ is $t$-quark in type 1, $b$-quark in types 2 and 3 and $\tau$-lepton in type 4.}
	\label{feynmanDiagram}
\end{figure}

The signal process under study is $e^{-}e^{+}\rightarrow Z^* \rightarrow HA$ in the four fermion final state as shown in Fig. \ref{feynmanDiagram} with the second vertex set to 
\begin{equation}
	g_{{HAZ}}=\frac{-g\sin(\beta-\alpha)}{2\cos\theta_W}
\end{equation}
which is maximum at the alignment limit. The light Higgs production, i.e., $e^{-}e^{+}\rightarrow Z^* \rightarrow hA$ involves $g_{hAZ}\propto \cos(\beta-\alpha)$ and vanishes due to suppression of  the vertex under the same circumstances.

As for the signal cross section, one can start from the usual differential form of
\begin{eqnarray}\label{sigma1}
	\dfrac{d\sigma}{d\Omega}=\frac{1}{64{\pi}^2 s}\dfrac{p_{f}}{p_{i}} \left \langle |M_{fi}|^{2} \right \rangle  
\end{eqnarray}
where in the center of mass frame, $p_i=\sqrt{s}/2$ and $p_f=\frac{1}{2\sqrt{s}}\sqrt{\lambda(s,m_H^2,m_A^2)}$ with $\lambda$ being the K\"{a}ll\'{e}n funtion defined as $\lambda(\alpha,\beta,\gamma)=\alpha^2+\beta^2+\gamma^2-2\alpha\beta-2\alpha\gamma-2\beta\gamma$. In this case one can write $\lambda$ as $[s-(m_A+m_H)^2][s-(m_A-m_H)^2]$. The matrix element consists of the $Z$ boson propagator, vertex couplings and the electroweak coefficients ($g^e_V\simeq-0.04$ and $g^e_A=-0.5$ according to particle data group \cite{pdg}) which lead to the following form for the integrated total cross section as a function of the collider center of mass energy $\sqrt{s}$ and the Higgs boson masses $m_H$ and $m_A$:
\begin{eqnarray}\label{sigma}
	\sigma = \dfrac{\pi \alpha^{2}(g_{V}^{2}+g_{A}^{2})[[s-(m_{A}+m_{H})^{2}][s-(m_{A}-m_{H})^{2}]]^{\frac{3}{2}} 
	}{48s^{2} \sin^{4}\theta_{W} \cos^{4}\theta_{W}[(s-m_{Z}^{2})^{2}+m_{Z}^{2}\Gamma_{Z}^{2}]}
\end{eqnarray}
According to the above form of Eq. \ref{sigma}, the cross section is suppressed when the sum of the Higgs boson masses approaches the center of mass energy, receives enhancement at the $Z$ boson pole mass, and prefers equal masses for the Higgs bosons due to the last term in the numerator. As the result a symmetric distribution is expected in the Higgs boson masses space. 

\begin{figure}[h]
	\includegraphics[width=\linewidth]{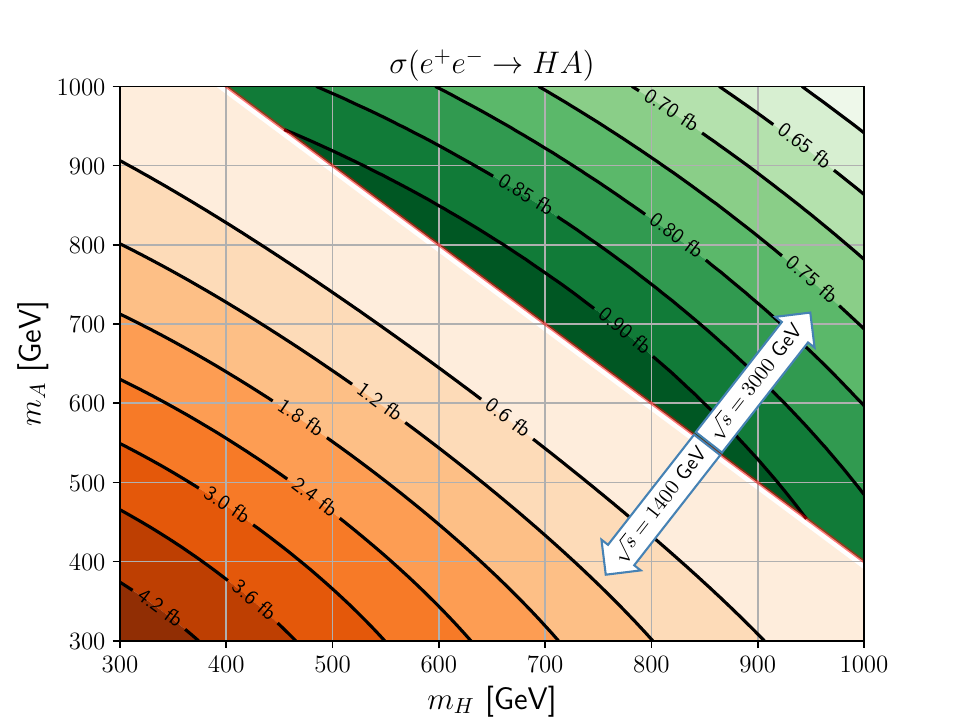}
	\caption{The signal cross section as a function of the Higgs boson masses for the two center of mass energies of 1400 and 3000 GeV, obtained using \texttt{WHIZARD} with unpolarized beams. }
	\label{splot}
\end{figure}

The tree level values predicted in Eq. \ref{sigma} are confirmed by \texttt{WHIZARD 3.1.2} \cite{whizard1,whizard2}. As an example, with $\sqrt{s}=1400$ GeV, $m_H=400$ GeV and $m_A=600$ GeV, Eq. \ref{sigma} yields $\sigma=2.2~fb$, to be compared with \texttt{WHIZARD} result of $\sigma=2.1~fb$ with unpolarized beams. Including a polarization of $80\%(30\%)$ for $e^-(e^+)$ enhances the above value to $3.1~fb$. A full scan over the Higgs boson masses in the two regions of $\sqrt{s}=1400$ and $3000$ GeV results in Fig. \ref{splot} where different color styles are used to distinguish the two regions of center of mass energies.     

The Higgs boson decay to fermion pairs can also be calculated using the square of the matrix element which, keeping the fermion masses, becomes:
\begin{eqnarray}\label{4}
	\sum_{\tn{spins}} |M_{H/A \to f\bar{f}}|^2 = 4\sqrt{2}G_Fm_f^2(\rho^f)^2 [(p_{f}.p_{\bar{f}}) \mp m_{f}^{2}]
\end{eqnarray}
where the subtraction(sum) on the right hand side occurs for CP-even(CP-odd) Higgs bosons respectively. The four momentum product is related to the particle masses through $s=m_{H/A}^2=(p_f+p_{\bar{f}})^2=2m_f^2+2p_f.p_{\bar{f}}$. As a result, CP-even and CP-odd Higgs bosons acquire different decay rates through their matrix elements:
\begin{equation}
	\langle |M^2| \rangle = 2\sqrt{2}G_Fm_f^2(\rho^f)^2 \times \Biggl \{ \begin{array}{ll}
		 m_H^2-4m_f^2 & $,H$  \\
		 m_A^2 & $,A$\\
		\end{array}
\end{equation} 
which result in (\cite{H2ff,A2ff}):
\begin{eqnarray}\label{decay}
	\Gamma (H/A \rightarrow f\bar{f})=\dfrac{3\sqrt{2}G_F m_{H/A} m_{f}^{2}(\rho^f)^2}{8\pi} \times \nonumber \\
	\Biggl \{ \begin{array}{ll}
		(1-\frac{4m_f^2}{m_H^2})^{3/2} & $,H$ \\
		(1-\frac{4m_f^2}{m_A^2})^{1/2} & $,A$\\
		\end{array}
\end{eqnarray}
With equal Higgs boson masses and couplings, additional factor of $1-4m_f^2/m_H^2$ in CP-even $H$ decay with respect to the CP-odd $A$, leads to suppression of $H \to t\bar{t}$ compared to $A\to t\bar{t}$ as can be seen in all model types. The consequence of this difference is that $A\to \tau\tau$ is less than $H\to \tau\tau$ above the kinematic threshold of decay to top quark pair in type 4. This phenomenon is most relevant when $m_{H/A}\sim 2m_f$, otherwise the above term of $1-4m_f^2/m_{H/A}^2$ tends to unity and the difference between the rates of $H \to t\bar{t}$ and $A \to t\bar{t}$ becomes smaller as $m_{H/A}$ tends to 1 TeV.  

\begin{figure}[h!]
	\includegraphics[width=\linewidth]{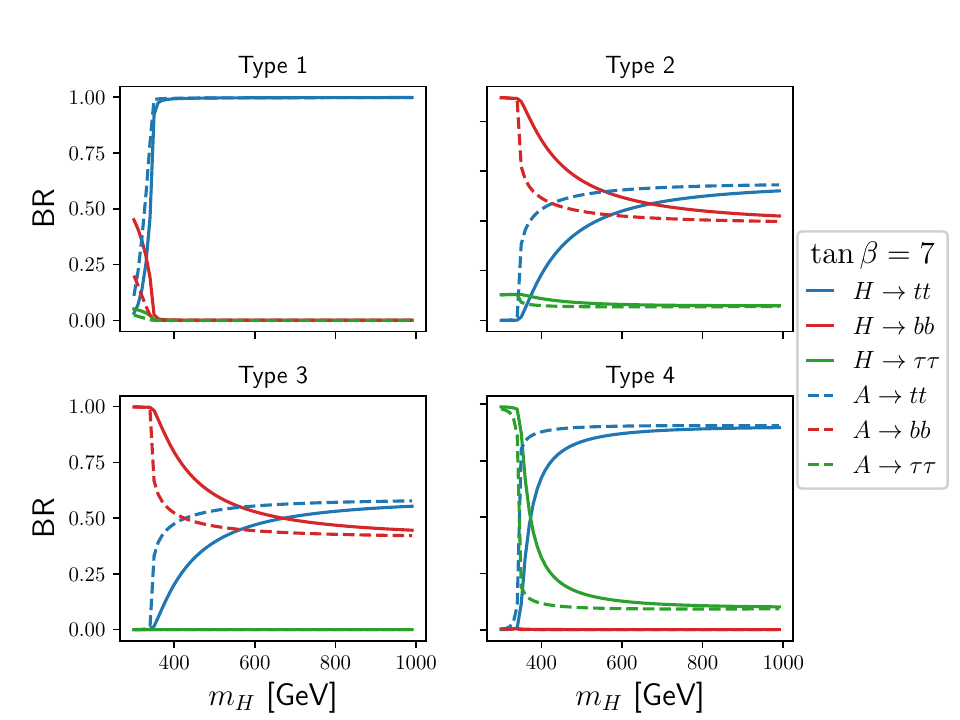}
	\caption{Branching ratio of Higgs boson decays to the main channels at $\tan\beta=7$. All heavy Higgs boson masses are assumed to be equal in this plot.}
	\label{BRs7}
\end{figure}

\begin{figure}[h!]
	\includegraphics[width=\linewidth]{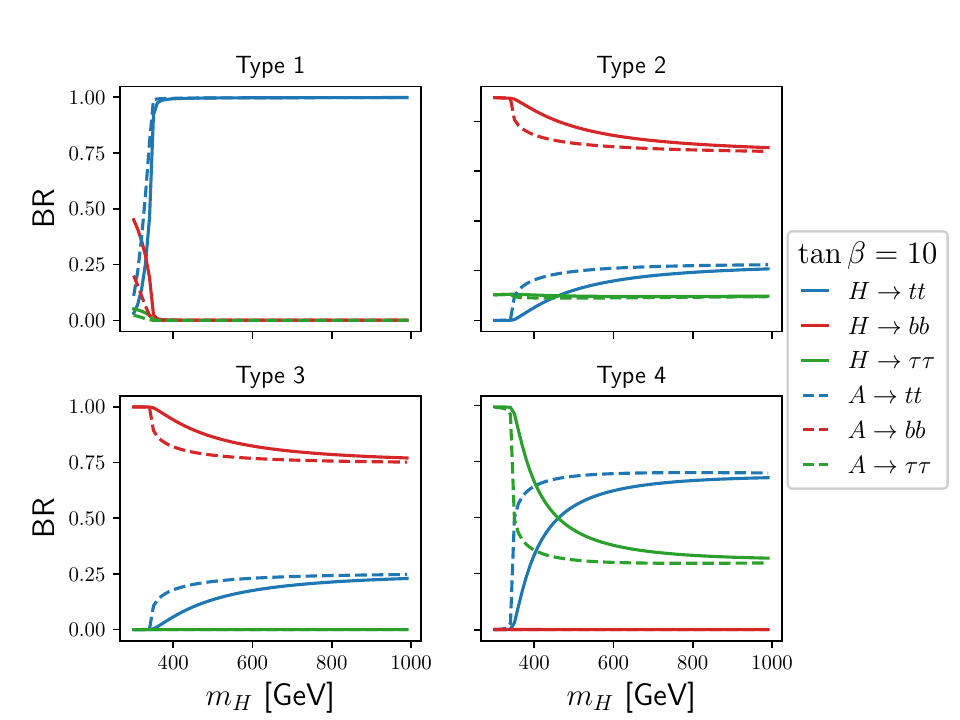}
	\caption{The same as Fig. \ref{BRs7} but with $\tan\beta=10$.}
	\label{BRs10}
\end{figure}

In order to visualize the above statements and decide on analysis channels, the Higgs boson decay rates are obtained using \texttt{2HDMC 1.8.0} \cite{2hdmc1,2hdmc2}. Since high $\tan\beta$ values are excluded in most model types, we start from $\tan\beta=7$ and 10 inspired from the solutions of $m_t \cot\beta = m_b \tan\beta$ and $m_t \cot\beta = m_\tau \tan\beta$ to obtain open regions for analysis. These $\tan\beta$ values are turning points above which $H\to bb/\tau\tau$ in types 2, 3, and 4 start to overwhelm $H\to t\bar{t}$ when $m_{H/A}>2m_t$. In type 4, the above argument is only valid for moderate masses up to 500 GeV while for heavier Higgs bosons, higher $\tan\beta$ values are needed to strengthen $H/A \to \tau\tau$. In type 1, regardless of $\tan\beta$, the dominant channel is $H\to t\bar{t}$ when kinematically allowed. Figures \ref{BRs7} and \ref{BRs10} show branching ratio of decays for both heavy neutral Higgs bosons in each type of the model at $\tan\beta=7$ and 10. 
Based on these figures the following final states are considered for each type:\\
$\bullet$\textbf{type 1}: $H/A\to b\bar{b}$ well below the top quark pair production threshold, i.e., $m_{H/A}<2m_t$ and $H/A\to t\bar{t}$ near or above the threshold. Since the current analysis deals with $m_{H/A}\geq 300$ GeV, the latter is chosen.\\
$\bullet$\textbf{type 2}: is actually highly suppressed by experimental searches for 2HDM and MSSM. A choice of $H/A\to b\bar{b}$ or $H/A\to \tau\tau$ is fine. If $H/A\to t\bar{t}$ is sizable (which is the case when $m_t \cot\beta \simeq m_b \tan\beta$), there is still preference in favor of $H/A\to b\bar{b}$ due to the lower final state particle multiplicity\\
$\bullet$\textbf{type 3}: $H/A\to b\bar{b}$ is dominant as long as high $\tan\beta$ values are considered. The same argument as in type 2 is applied to $H/A\to t\bar{t}$.\\
$\bullet$\textbf{type 4}: $H/A\to \tau\tau$ is dominant as long as high $\tan\beta$ values are considered. In this case also $H/A\to t\bar{t}$ can be sizable (when $m_t \cot\beta \simeq m_{\tau} \tan\beta$). However, due to the same reasons as mentioned in type 2, $H/A\to \tau\tau$ is preferred.

The experimental exclusion regions are obtained using \texttt{HiggsTools-1} \cite{higgstools} which is a collection of \texttt{HiggsBounds-6} \cite{hb1,hb2,hb3,hb4,hb5}, \texttt{HiggsSignals-3} \cite{hs1,hs2,hs3} and the new code \texttt{HiggsPredictions-1}. At the moment, results of 258 analyses from LEP to LHC are included in the code database. The input data for obtaining excluded regions are the scaled effective Higgs-fermion and Higgs-gauge couplings relative to SM and the Higgs boson decay rates for non-fermionic channels like $H\to hh$ and $A\to ZH$ which are taken (borrowed) from \texttt{2HDMC}.

\begin{figure}[h!]
	\includegraphics[width=\linewidth]{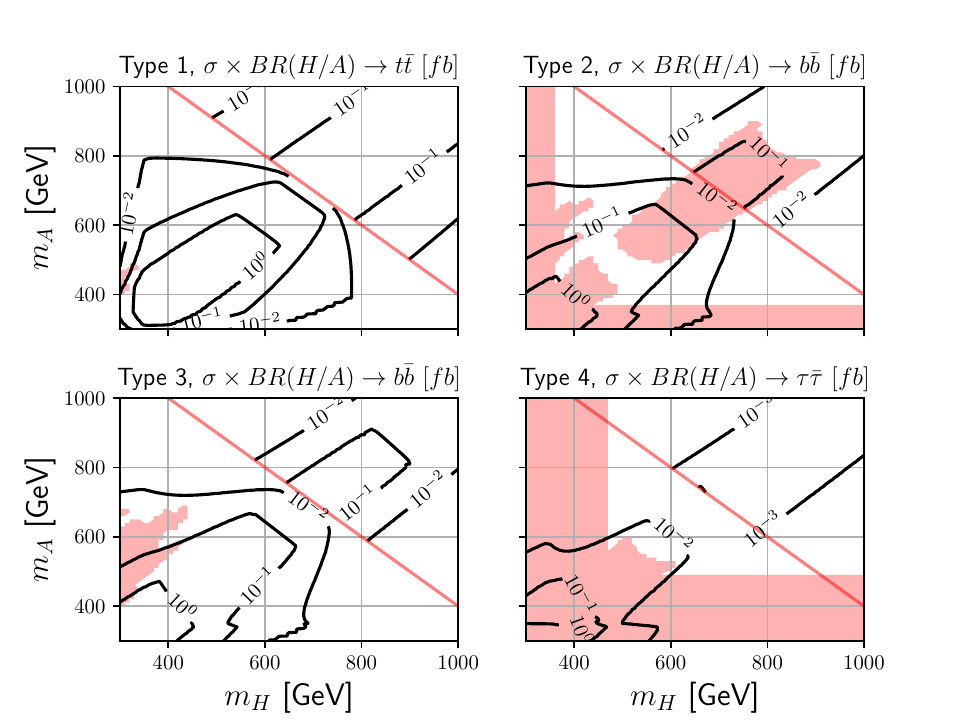}
	\caption{The signal cross section times branching ratio of decay to relevant final states in each type at $\tan\beta=7$. The current excluded regions at 95$\%$ CL are shown in solid red.}
	\label{sxbr7}
\end{figure}

\begin{figure}[h!]
	\includegraphics[width=\linewidth]{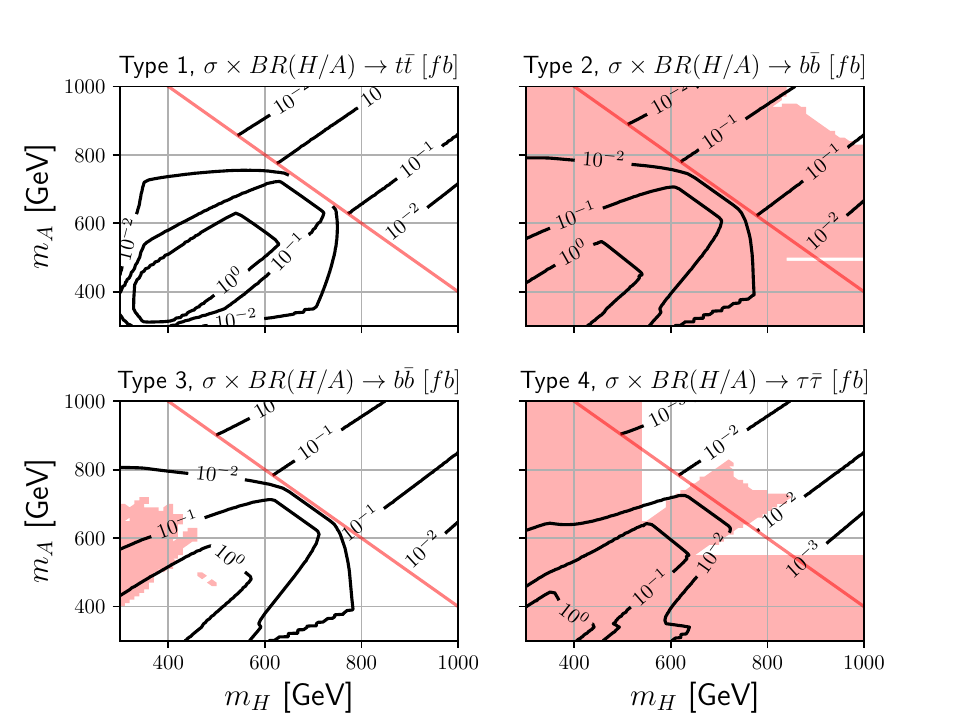}
	\caption{The signal cross section times branching ratios with the same descriptions as in Fg. \ref{sxbr7} but with $\tan\beta=10$.}
	\label{sxbr10}
\end{figure}

Figures \ref{sxbr7} and \ref{sxbr10} show the resulting excluded regions for $\tan\beta=7$ and 10 together with the signal cross section times branching ratio of decay to the relevant final state in each type. The red line indicates the border of two scenarios of the collider operation at $\sqrt{s}=1400$ and 3000 GeV. The excluded regions in type 1 (at $\tan\beta=7$) and type 3 at $m_H<400$ GeV refer to the LHC analysis of Higgs boson conversion reported in \cite{T3_2}. In type 3, at $\tan\beta=10$ there are two points of $(m_H,m_A)=(470,490)$ and $(490,470)$ GeV excluded by LHC \cite{T3_1}. The type 2 is excluded mainly by the two LHC analyses reported in \cite{T3_2,T2}. The type 4 excluded regions refer to \cite{T2} and \cite{T4_1,T4_2}. In types 3 and 4, the higher the $\tan\beta$, the more signal sensitivity due to the higher branching ratio of Higgs boson decay. Therefore $\tan\beta=10$ is chosen for the rest of the analysis. At this value, type 2 is already excluded up to near 1 TeV. The LHC exclusion of type 2 parameter space is in agreement with prediction reported in \cite{DJMSSM1} where it was shown that MSSM Higgs boson masses up to 1 TeV at $\tan\beta=10$ can be excluded at HL-LHC integrated luminosity of 300 $fb^{-1}$. Higher masses up to several TeV were analyzed in \cite{DJMSSM2} for a hadron collider operating at 100 TeV (FCC-hh). These results and current LHC experimental results shown in Figs. \ref{sxbr7} and \ref{sxbr10} (based on \cite{T3_2,T3_1,T2,T4_1,T4_2}) show that the current open regions of the parameter space are reasonable choices for CLIC studies.        
\section{Signal and Background generation}
The main background events are electroweak boson pair production $ZZ$ (including $\gamma$) and top quark pair production $t\bar{t}$. The single boson production (Drell-Yan) does not contribute as it contains less number of jets needed to fully reconstruct events. The $W^+W^-$ background contribution is also negligible due to the presence of the light jets in the final state and low fake rate. 

Both signal and background processes are generated starting from \texttt{WHIZARD} which produces \texttt{LHEF} files containing the hard scattering \cite{lhef}. These files are used by \texttt{PYTHIA 8.3.09} \cite{pythia} which is called through \texttt{DELPHES} \cite{delphes1,delphes2,delphes3} for fast detector simulation. The jet reconstruction is performed using Valencia algorithm \cite{VLC} as proposed by the CLIC collaboration \cite{CLICdp} due to the robust performance in the non-negligible background environment even better than the $k_T$ algorithm \cite{kt}. 

\begin{figure}[h!]
	\includegraphics[width=\linewidth]{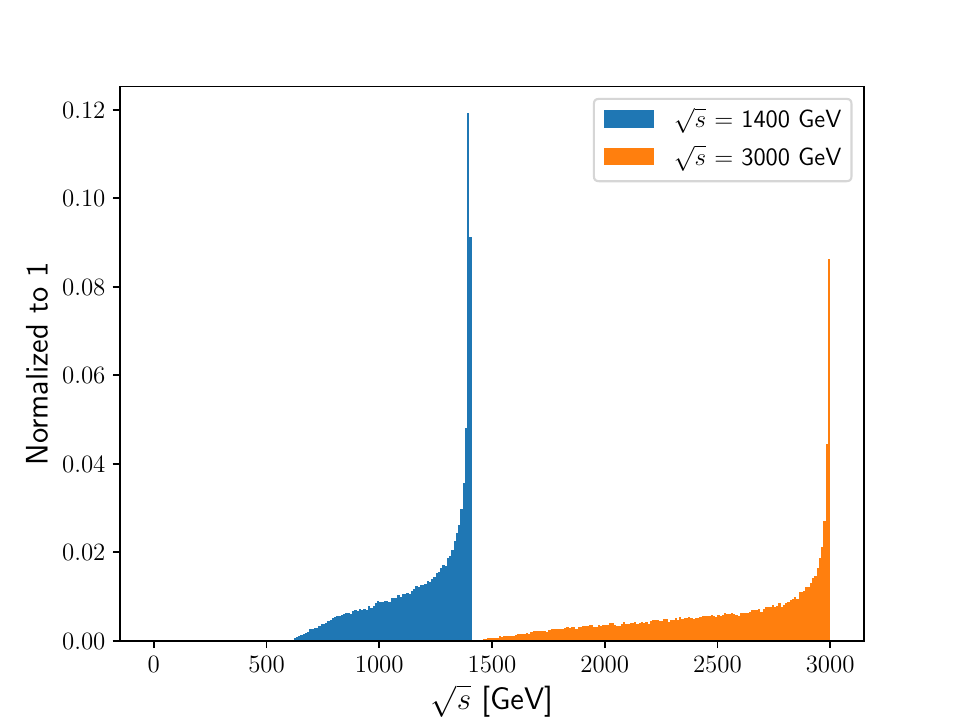}
	\caption{Beam spectrum at $\sqrt{s}=1400$ and 3000 GeV, simulated by \texttt{CIRCE2} and used for event generation by \texttt{WHIZARD}.}
	\label{beamspectrum}
\end{figure}

The \texttt{WHIZARD} package performs bremsstrahlung and initial state radiation from the beam particles using \texttt{CIRCE2} code \cite{circe2} resulting in the beam spectrum as shown in Fig. \ref{beamspectrum}. The beamsstrahlung not only reduces the beam energy but also results in radiated photon collisions as the so called overlay background \cite{overlay}. These events have been shown to be reducing the jet energy resolution in preliminary studies and their effects are minimally included as a source of jet energy smearing applied after the jet reconstruction by \texttt{DELPHES}. The values implemented for the two scenarios of 1400 and 3000 GeV are shown in Tab. \ref{overlaytab}.  
\begin{table}[h]
		\centering
		\begin{tabular}{|c|c|c|}\hline
			$\sqrt{s}$ [GeV]& 1400 & 3000 \\
			\hline 
			$|\eta|<0.76$ & 1$\%$ & 2$\%$ \\
			\hline
			$|\eta|\geq 0.76$ & 5$\%$ & 10$\%$ \\
            \hline
            \end{tabular}
        \caption{The jet energy smearing in percent applied on jets in the two regions of pseudorapidity to mimic the $\gamma\gamma \to \textnormal{hadrons}$ overlay background.}
        \label{overlaytab}
        \end{table}
\section{Benchmark points}
Figure \ref{sxbr10} shows that the signal rates are higher around the diagonal line of degenerate masses due to the cross section symmetry. Therefore the set of benchmark points for the analysis are divided into two categories of degenerate and non-degenerate masses. The set of $m_H=m_A$ points are chosen with increments of 100 GeV starting from 300 to 600 GeV for collider operation at $\sqrt{s}=1400$ GeV and 700 to 1000 GeV for the scenario of $\sqrt{s}=3000$ GeV. The non-degenerate points are chosen with the same CP-even heavy Higgs boson masses but with $m_A=m_H+50$ GeV for the two center of mass energies. These points are relevant to types 1 and 3. In type 4, $m_H$ starts from 800 GeV as lower masses are already excluded. 

The $Z_2$ soft breaking mass parameter $m_{12}$ is set through the relation $m_{12}^2=m_H^2\sin\beta\cos\beta$ which is suggested in \cite{m12} as a consequence of natural SM alignment of the model without requiring the decoupling or a fine tuning. For completeness, we recall other parameters of the model: $\tan\beta=10$ and $\sin(\beta-\alpha)=1$. 

Tables \ref{sigcs} and \ref{Bcs} show the signal and background cross sections as a numeric reference. It should be noted that the background processes are generated in fully hadronic final state to reduce the statistics. Therefore, $ZZ$ and $t\bar{t}$ backgrounds undergo $Z\to q\bar{q}$ and $W \to q\bar{q'}$ respectively with branching ratios of 0.69 and 0.68 \cite{pdg}. The $hZ$ process is also analyzed in the four $b$-jet final state in type 3 with BR($h\to b\bar{b}$)=0.55. In type 4, this background has no sizable contribution in the signal region.         
\begin{table}[h]
\centering
\begin{tabular}{|c|ccccc|}
\hline
$\sqrt{s}$ [GeV]& $m_H=m_A$= & 300 & 400 & 500 & 600\\
\multirow{3}{*}{1400}& $\sigma [fb]$ & 4.8 & 2.8 & 1.3 & 0.37 \\
\cline{2-6}
& $m_A=m_H+50=$ & 350 & 450 & 550 & 650 \\
& $\sigma [fb]$ & 4.3 & 2.4 & 1.1 & 0.22 \\
\hline
& $m_H=m_A$= & 700 & 800 & 900 & 1000\\
\multirow{3}{*}{3000}& $\sigma [fb]$ & 0.84 & 0.64 & 0.48 & 0.34 \\
\cline{2-6}
& $m_A=m_H+50=$ & 750 & 850 & 950 & 1050 \\
& $\sigma [fb]$ & 0.79 & 0.6 & 0.44 & 0.31 \\
\hline
\end{tabular}
\caption{Signal cross section at benchmark points.}
\label{sigcs}
\end{table}

\begin{table}[h!]
\centering
\begin{tabular}{|c|cccc|}
\hline
$\sqrt{s}$ && $ZZ$ & $t\bar{t}$ & $hZ$ \\
\cline{2-5}
1400 & $\sigma[fb]$& 142 & 145 & 13.5 \\
\cline{2-5}
3000 & $\sigma[fb]$& 58 & 53 & 4.7 \\
\hline
\end{tabular}
\caption{Background cross sections.}
\label{Bcs}
\end{table}
\section{Analysis, type 1}
The type 1 events are predominantly in the four top quark final state above the kinematic threshold. Events are therefore produced with high particle multiplicity if top quarks decay hadronically. With a normal cone size of $0.5$ one would expect to have three jets from each top quark in its fully hadronic final state resulting in a total of 12 jets. In this case the top tagging algorithm \cite{tt1,tt3,tt5,tt7} can be used to identify the highly boosted top quarks based on the jet substructure \cite{jss2,jss6,jss7,jss8}. There are several top tagging algorithms two of which are used in this analysis namely \texttt{HEPTopTagger2} \cite{HEPTopTagger1,HEPTopTagger2} and \texttt{JohnHopkins} \cite{JohnHopkins}. The external codes for these algorithms are linked to the \texttt{FastJetFinder} jet reconstruction module inside \texttt{DELPHES}. The performance of the algorithm is expected to increase with increasing Higgs boson masses thus producing highly boosted top quarks. The jet reconstruction cone is set to 0.8 with Cambridge-Aachen algorithm used for the jet reconstruction \cite{CA}. This choice of the jet cone gives in average four identified jets to be used as input by the top tagging algorithm. 

\begin{figure}[h]
	\includegraphics[width=\linewidth]{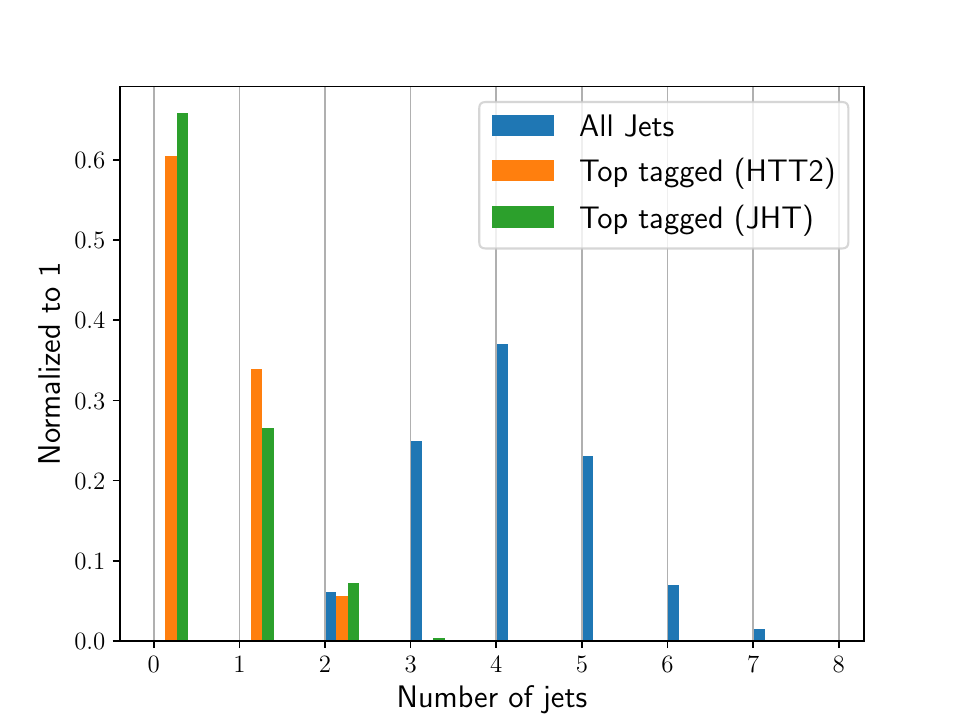}
	\caption{Number of identified top jets in the signal events at $\sqrt{s}=3000$ GeV using two top tagging algorithms \texttt{HEPTopTagger2} (labeled HTT2) and \texttt{JohnHopkins} (labeled JHT).}
	\label{jetmul_tt}
\end{figure}

The identified top jets from Higgs bosons with $m_H=m_A=700$ GeV at $\sqrt{s}=3000$ GeV result in the multiplicity distribution as shown in Fig. \ref{jetmul_tt}. As seen the jet multiplicity has a peak at four, while identified top jets have no sizable contribution in the four top bin for both algorithms. Therefore, with the choice of the top tagging algorithms examined here, the signal can not be reconstructed unless a detailed study of this type with more sophisticated ML based top tagging algorithms is performed.  
\section{Analysis, type 2}
Although the type 2 is very limited at $\tan\beta=10$ which is the base for the analysis, one point worthy to mention is that the signal topology and kinematics of the final state makes no difference between the types 2 and 3 with the same Higgs boson masses and decay channels. Therefore, taking into account branching ratio of Higgs bosons decays in two types, the signal sensitivity in type 3 can easily be normalized to get the corresponding results in type 2. 

As an example, at $\tan\beta=7$, two scenarios of $m_H=m_A=500$ GeV and $m_H=m_A=900$ GeV, which are not yet excluded, lead to the following BRs for the four $b$-jet final state as shown in Tab. \ref{BRs}. Due to the similarity of figures, one would expect to get similar results in type 2 by translating an analysis in the same final state in type 3.
\begin{table}[h]
	\centering
	\begin{tabular}{|c|c|cc|}
		\hline
		$m_H=m_A=$& &$H\to b\bar{b}$ & $A\to b\bar{b}$\\
		\multirow{2}{*}{500 GeV}& Type 2 & 0.49 & 0.36 \\
		\cline{2-4}
		&Type 3& 0.53 & 0.39 \\
        \hline
		$m_H=m_A=$& &$H\to b\bar{b}$ & $A\to b\bar{b}$\\
\multirow{2}{*}{900 GeV}& Type 2& 0.35 & 0.33 \\
\cline{2-4}
&Type 3 & 0.38 & 0.35 \\
\hline
	\end{tabular}
	\caption{Branching ratio of CP-even and CP-odd Higgs boson decay to $b\bar{b}$ in types 2 and 3 for the two points outside the current excluded area.}
	\label{BRs}
\end{table}
 
\section{Analysis, type 3}
\begin{figure}[h!]
	\centering
	\includegraphics[width=\linewidth]{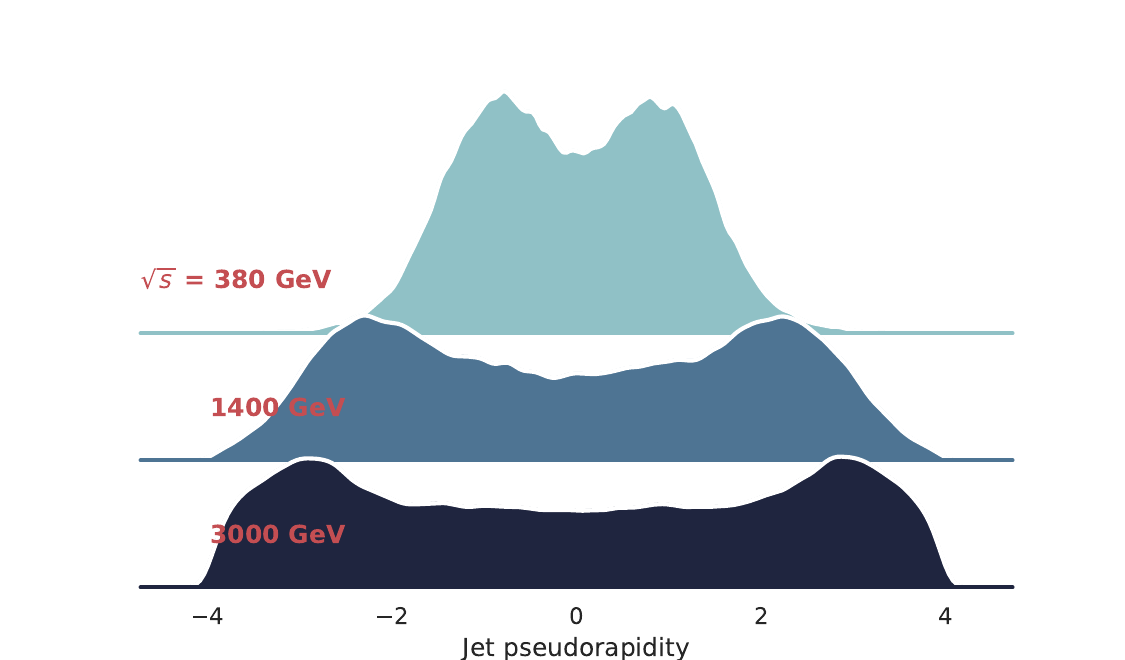}
	\caption{The jet pseudorapidity distribution in $ZZ$ background as a function of the collider center of mass energy.}
	\label{eta}
\end{figure}

The type 3 events are analyzed in the four $b$-jet final state using the $b$-tagging algorithm optimized for CLIC detector environment \cite{clicb}. The kinematic acceptance for the jet reconstruction is $p_T>10$ GeV and $|\eta|<2$ with pseudorapidity defined as $\eta=-\ln\tan\theta/2$ ($\theta$ is the polar angle with respect to the beam axis). 
The $ZZ$ background shows less efficiency of four jet reconstruction compared to other backgrounds due to the electroweak nature of these events and wide pseudorapidity distribution of the jets as shown in Fig. \ref{eta}. The kinematic acceptance has therefore lower efficiency of selection at higher center of mass energies for this background.

\begin{figure}[h]
	\includegraphics[width=0.9\linewidth]{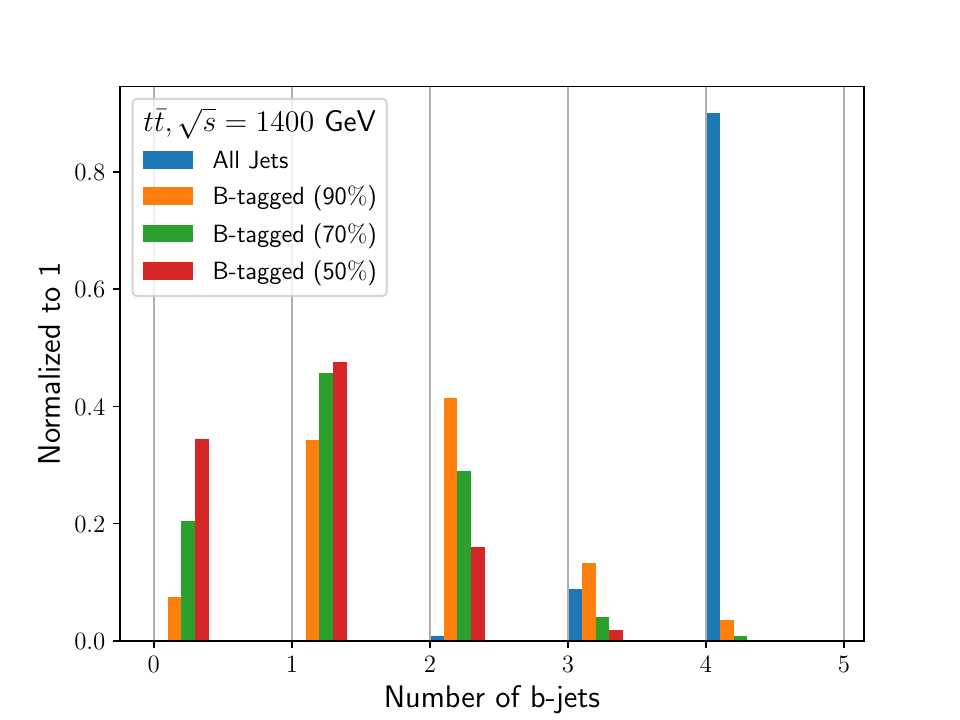}
	\caption{Number of identified $b$-jets with different $b$-tagging efficiencies in $t\bar{t}$ events at $\sqrt{s}=1400$ GeV.}
	\label{bmul_tt}
\end{figure}

\begin{figure}[h]
	\includegraphics[width=0.9\linewidth]{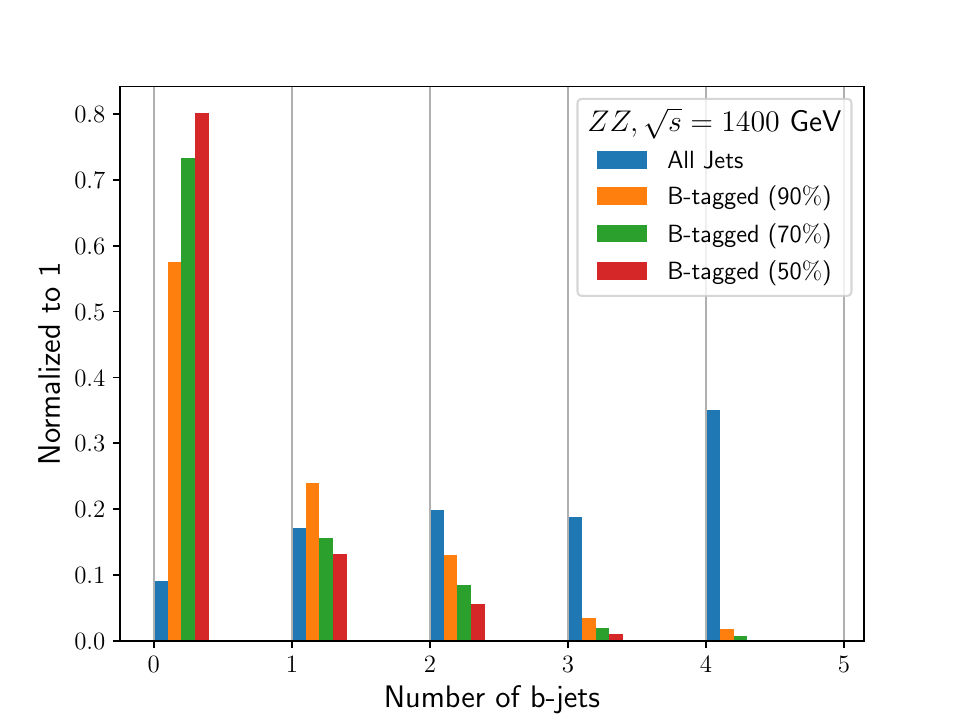}
	\caption{Number of identified $b$-jets with different $b$-tagging efficiencies in $ZZ$ events at $\sqrt{s}=1400$ GeV.}
	\label{bmul_ZZ}
\end{figure}

\begin{figure}[h]
	\includegraphics[width=0.9\linewidth]{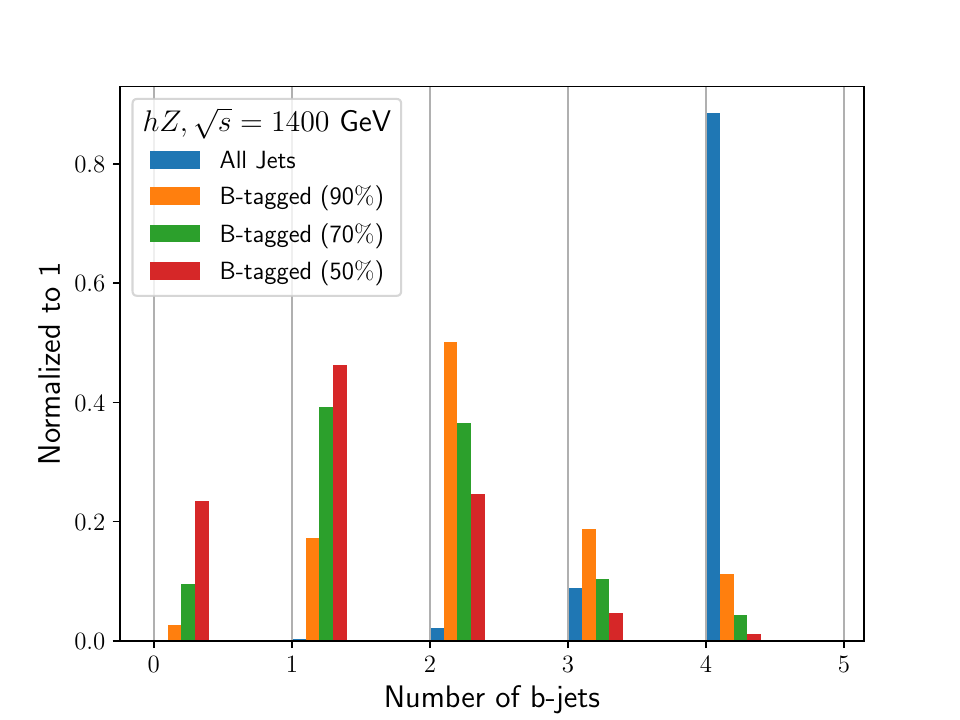}
	\caption{Number of identified $b$-jets with different $b$-tagging efficiencies in $hZ$ events at $\sqrt{s}=1400$ GeV.}
	\label{bmul_hZ}
\end{figure}

There are three scenarios of $b$-tagging with the efficiencies of $90\%,~70\%$ and $50\%$ including fake rate from gluons and $c$-jets as a function of the jet energy and pseudorapidity. These working points are examined for the main background samples, $t\bar{t}$, $ZZ$ and $hZ$ with the reconstructed jet multiplicities as well as those of $b$-jets shown in Figs. \ref{bmul_tt}, \ref{bmul_ZZ} and \ref{bmul_hZ}. 

The lower $b$-tagging efficiency, the higher suppression of the background. However, even the loose $b$-tagging scenario with 90$\%$ efficiency allows a very small fraction of background events to fill the four $b$-jet bin and is thus chosen for the rest of the analysis.       

\begin{figure}[h!]
	\centering
	\includegraphics[width=\linewidth]{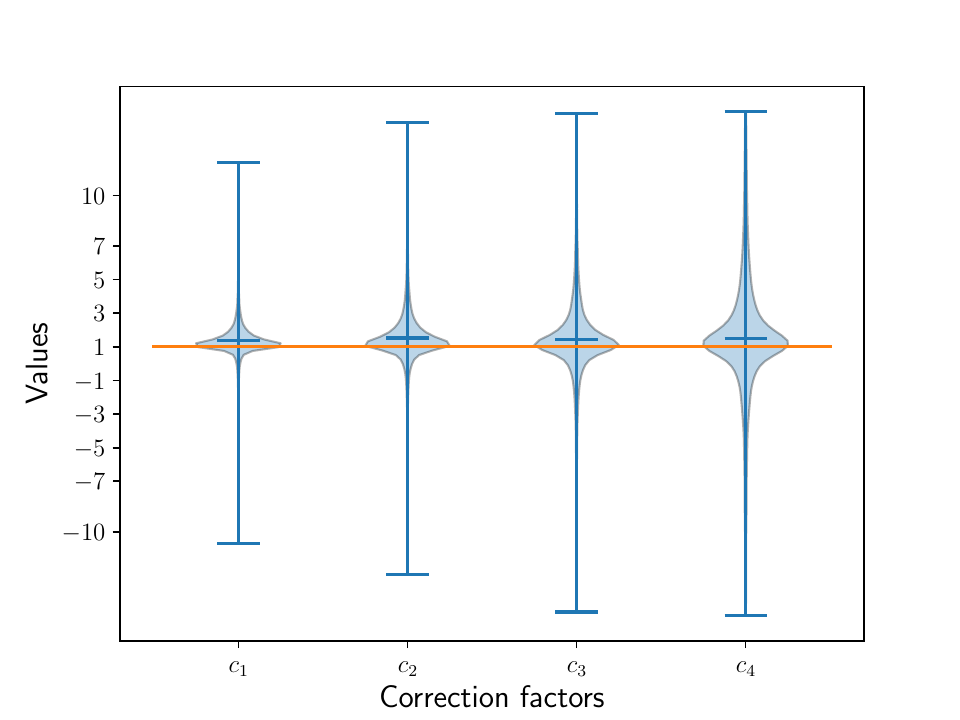}
	\caption{Distribution of the correction factors obtained by solving Eqs. \ref{emc}.}
	\label{ci}
\end{figure}

In the next step, a kinematic fit is applied by imposing energy and momentum conservation on events with four reconstructed $b$-jets as proposed in LEP (ALEPH) analysis \cite{ALEPH}. The idea is to apply correction factors to the $b$-jets so that the set of four-momentum conservation equations (Eqs. \ref{emc}) are satisfied:
\begin{eqnarray}
	\sum_{j=1}^4c_j\vec{p}_j&=&0\nonumber\\
	\sum_{j=1}^4c_jE_j&=&\sqrt{s}
	\label{emc}
\end{eqnarray}
Despite the uncertainty in the center of mass energy due to the beam spectrum, correction factors yield reasonable results as will be seen soon. Figure \ref{ci} shows the distribution of the correction factors $c_j$ with their average values close to (but slightly above) unity for an example benchmark point of $m_H=m_A=300$ GeV at $\sqrt{s}=1400$ GeV.  

The $b$-jets are sorted in terms of their energies before correction. The wider distributions for softer jets is due to the more uncertainty in the jet reconstruction with lower energy. Since all four-momentum components of each $b$-jet is to be multiplied by the corresponding correction factor, we require that all $c_j$ with $j=1$ to 4 are positive to avoid negative energies after correction. The energy sorting is repeated after correction and $b$-jets are labeled as $b_i$ with $i=1$ to 4 denoting the hardest to the softest $b$-jet in the list. 

\begin{figure*}
	\centering
	\begin{subfigure}{.5\textwidth}
		\centering
		\includegraphics[width=\linewidth]{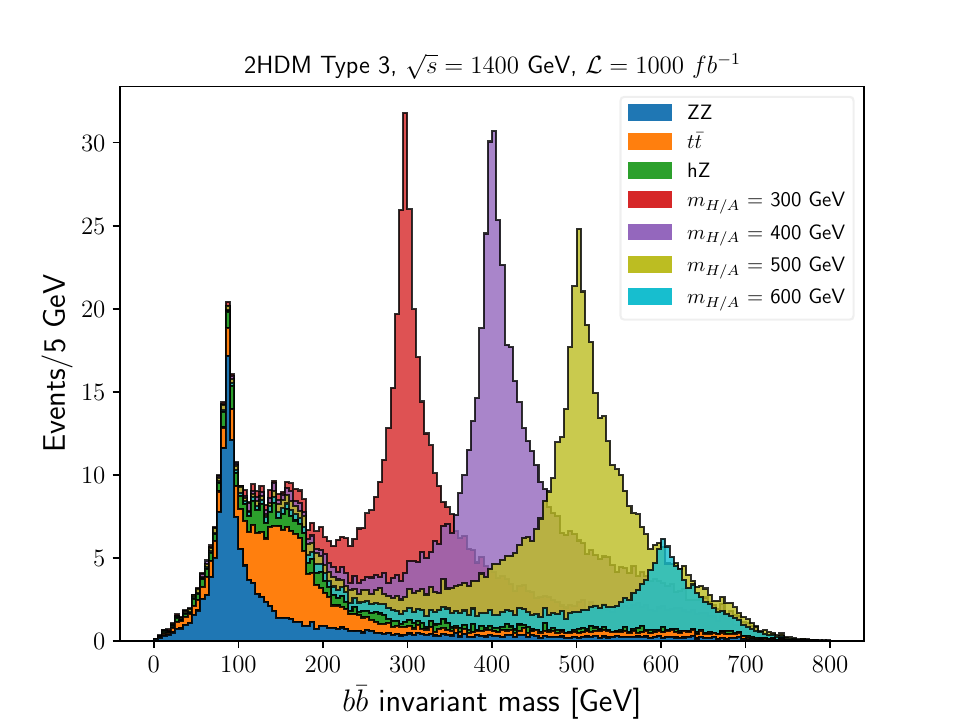}
		\caption{Equal masses}
		\label{EM1400}
	\end{subfigure}%
	\begin{subfigure}{.5\textwidth}
		\centering
		\includegraphics[width=\linewidth]{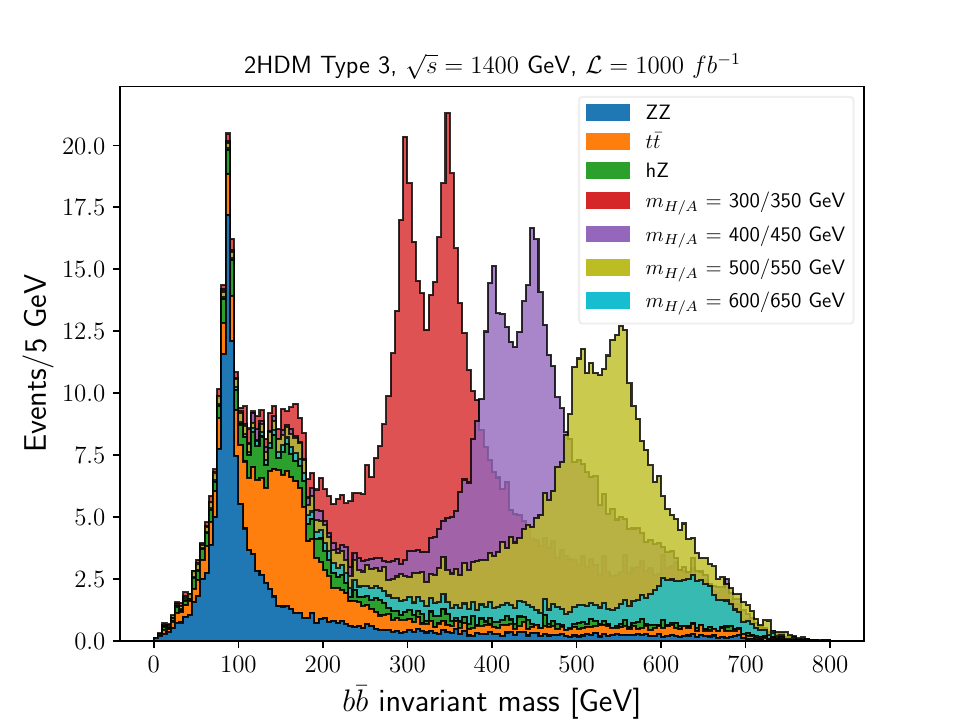}
		\caption{Different masses}
		\label{DM1400}
	\end{subfigure}
	\caption{The b-jet pair invariant mass distributions in signal and background events for type 3 with $\sqrt{s}=1400$ GeV. The signals containing $m_H=m_A=300$ and 400 GeV are divided by 5 and 2 and those with $m_A=m_H+50=350$ and 450 are divided by 3 and 2 to fit the window.}
\end{figure*}

\begin{figure*}
	\centering
	\begin{subfigure}{.5\textwidth}
		\centering
		\includegraphics[width=\linewidth]{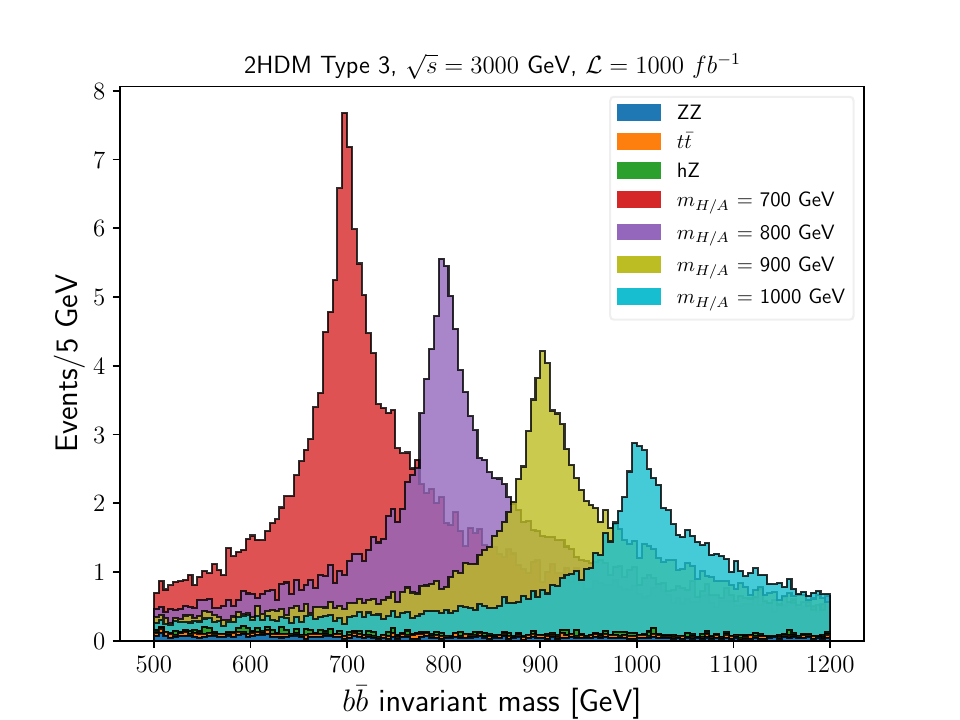}
		\caption{Equal masses}
		\label{EM3000}
	\end{subfigure}%
	\begin{subfigure}{.5\textwidth}
		\centering
		\includegraphics[width=\linewidth]{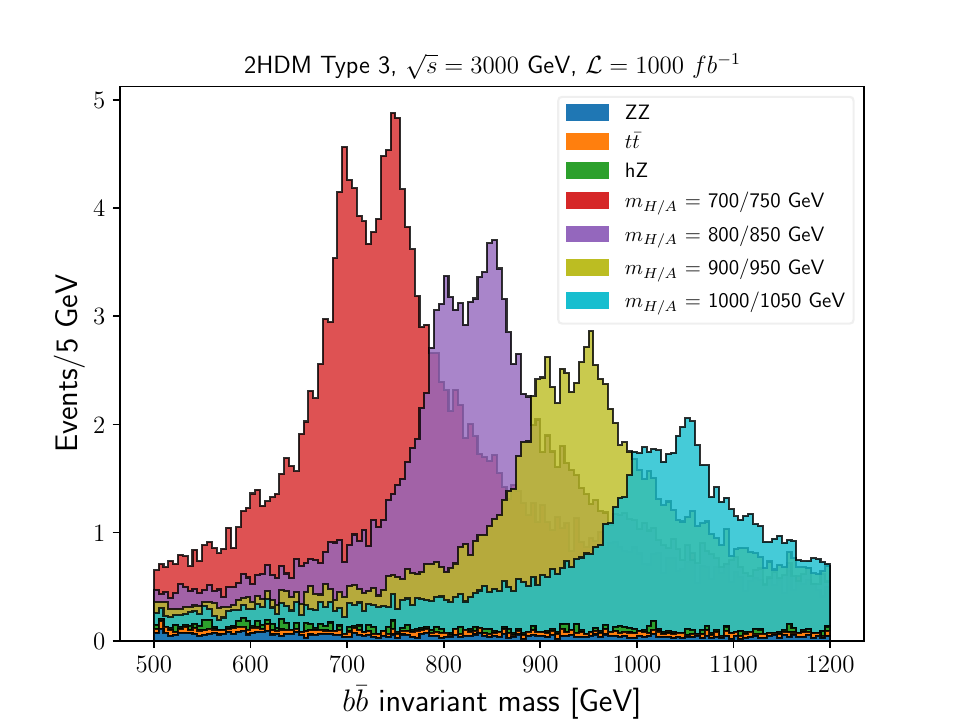}
		\caption{Different masses}
		\label{DM3000}
	\end{subfigure}
	\caption{The b-jet pair invariant mass distributions in signal and background events for type 3 with $\sqrt{s}=3000$ GeV.}
\end{figure*}

In the next step, $b$-jet pairing is performed as described in \cite{HE} using invariant masses of $b_2b_3$ and $b_1b_4$ together in one histogram. Although $b_2b_3$ pairs have proved to result in better invariant mass distributions for the Higgs bosons, the two pairs are used to get double statistics. Figures \ref{EM1400} (equal masses) and \ref{DM1400} (different masses) show the reconstructed Higgs boson invariant masses on top of the background at $\sqrt{s}=1400$ GeV. Results of higher masses at $\sqrt{s}=3000$ GeV are shown in Figs. \ref{EM3000} and \ref{DM3000}. Signals are stacked on the background but not on each other. As seen from Figs. \ref{DM1400} and \ref{DM3000} the two Higgs bosons can be distinguished if their masses are different. 

In obtaining these distributions, the charged track momentum smearing and ECAL/HCAL energy resolution are included using the standard CLIC detector card implemented in \texttt{DELPHES}. 

\begin{figure}[h!]
	\centering
	\includegraphics[width=\linewidth]{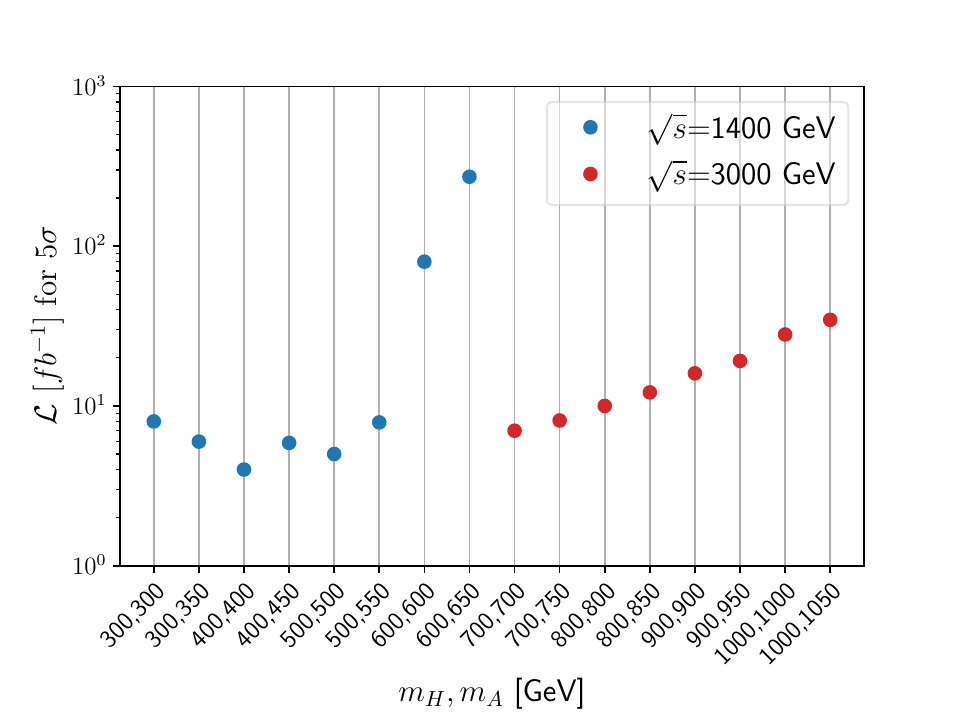}
	\caption{Integrated luminosity $\mathcal{L}$ needed for $5\sigma$ discovery in 2HDM type 3.}
	\label{L3}
\end{figure}

A mass window optimization is performed to get the highest signal significance, $S/\sqrt{B}$, and values exceeding $5\sigma$ are obtained in all benchmark points at integrated luminosity of $1000~fb^{-1}$. Therefore, final results are presented in terms of integrated luminosity needed for $5\sigma$ discovery as shown in Fig. \ref{L3}. The lowest signal sensitivity for $m_H=600$ GeV is due to $m_H+m_A$ getting close to the center of mass energy thus saturating the phase space. As the result, the cross section is small and more data is needed for the signal observation.          	

\section{Analysis, type 4}
The type 4 or the so-called lepton-specific 2HDM prefers leptonic decays of the Higgs bosons with Higgs-fermion couplings proportional to $\tan\beta$. At $\tan\beta=10$, $m_t\cot\beta \sim m_{\tau}\tan\beta$ and the two decay modes compete. The $A \to \tau\tau$ is slightly smaller than $H \to \tau\tau$ due to what was said in Eq. \ref{decay}. At higher $\tan\beta$, $H/A \to \tau\tau$ supersedes $H/A \to t\bar{t}$ and the signal rate is higher. Therefore we take $\tan\beta=10$ as the bottom line above which higher S/B is expected. 

The $\tau$-tagging algorithm in the hadronic final state is performed with efficiencies depending on $p_T$ and $\eta$ of the $\tau$-jet. The fake rate is assumed to be $3\%$ with the $\tau$-tagging efficiency around 60$\%$ for $p_T>50$ GeV \cite{clictau}. 

\begin{figure*}
	\centering
	\begin{subfigure}{.5\textwidth}
		\centering
		\includegraphics[width=\linewidth]{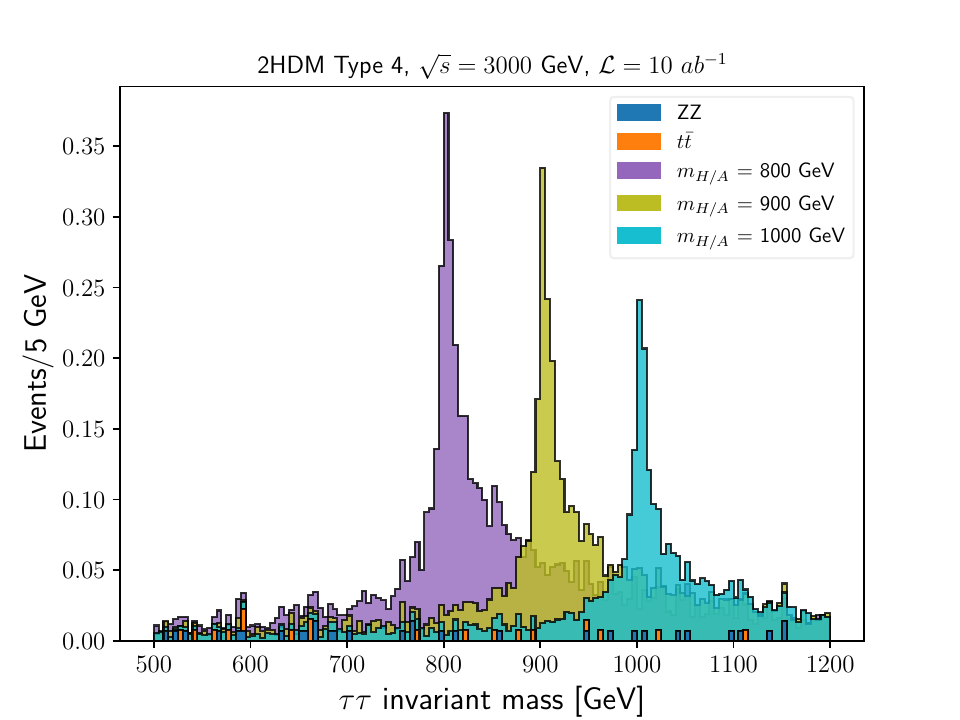}
		\caption{Equal masses}
		\label{EMtau}
	\end{subfigure}%
	\begin{subfigure}{.5\textwidth}
		\centering
		\includegraphics[width=\linewidth]{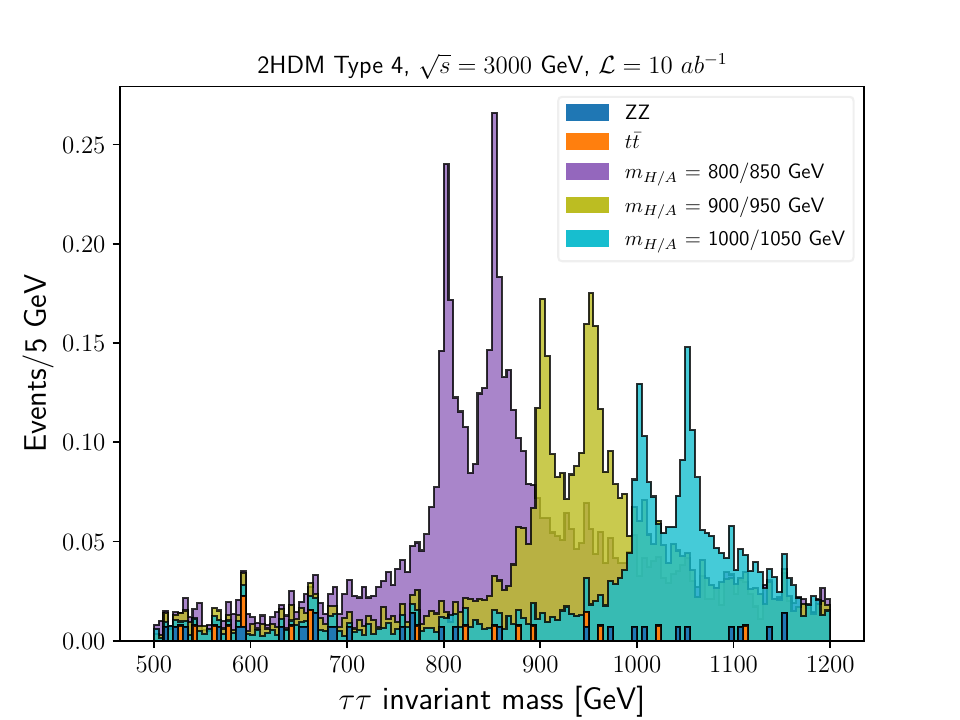}
		\caption{Different masses}
		\label{DMtau}
	\end{subfigure}
	\caption{The $\tau$-jet pair invariant mass distributions in signal and background events for type 4 with $\sqrt{s}=3$ TeV.}
\end{figure*}

As seen from Fig. \ref{sxbr10}, the Higgs boson masses below 800 GeV are already excluded in type 4 at $\tan\beta=10$, therefore three scenarios of $m_H=m_A=800,~900,~1000$ GeV and $m_A=m_H+50=850,~950,~1050$ GeV are examined in the four $\tau$-jet final state. The kinematic selection and procedure for $\tau$-jet pairing is the same as what described in the previous section. There is small background left in the signal region in this case and the signal lies on an almost background free region as shown in Figs. \ref{EMtau} and \ref{DMtau}. 

\begin{figure}[h!]
	\centering
	\includegraphics[width=\linewidth]{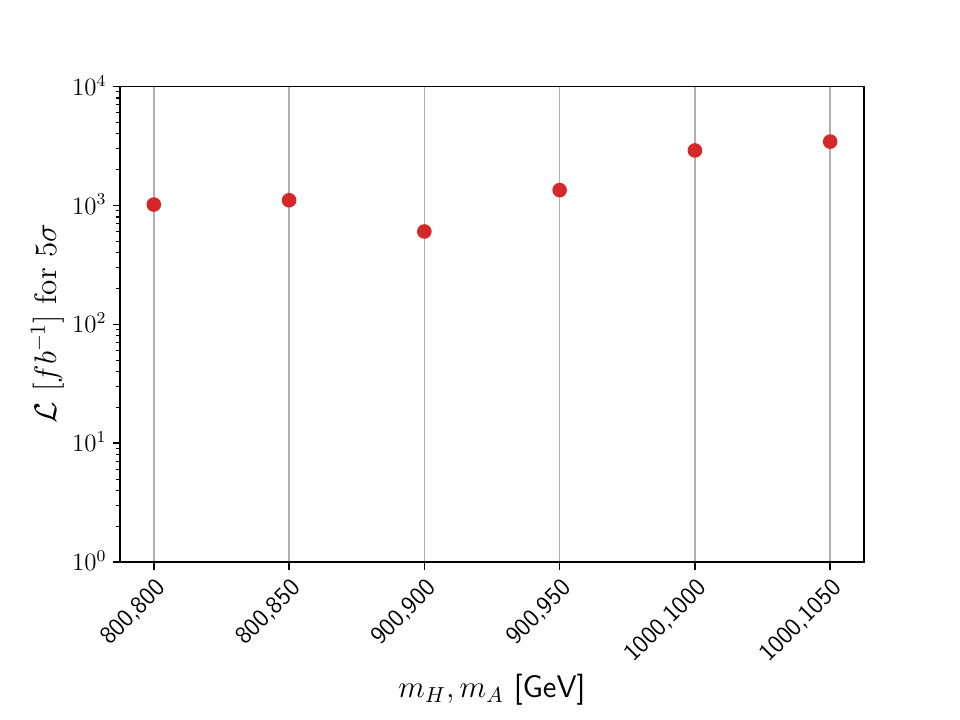}
	\caption{Integrated luminosity $\mathcal{L}$ needed for $5\sigma$ discovery in 2HDM type 4.}
	\label{L4}
\end{figure}

The signal distributions are sharper than the case of four $b$-jet final state in type 3. However, due to the lower signal rate and $\tau$ tagging efficiencies, more data or integrated luminosity is needed for the $5\sigma$ discovery compared to the case of type 3. Figure \ref{L4} shows integrated luminosity in $fb^{-1}$ needed for $5\sigma$ discovery. 

\section{Conclusions}
The heavy neutral 2HDM Higgs boson pair production was studied in the mass range of 300 to 1000 GeV at two stages of the CLIC operation, i.e., $\sqrt{s}=1400$ and 3000 GeV. The signal was analyzed in the relevant final states of each 2HDM type in the regions of the parameter space not yet excluded by the current LHC 8 and 13 TeV analyses. 

The type 1 with its dominant final state of four top quarks shows poor top tagging performance with the two exploited algorithms. A more sophisticated analysis using machine learning algorithms is needed to obtain a reasonable signal in this type. While type 2 with its similar decay rates to type 3 at $\tan\beta\simeq10$ is almost fully excluded, type 3 shows a promising signal on top of the background with distinct signals for CP-even and CP-odd Higgs bosons in scenarios with different masses. The type 4, shows a sharper but with less statistics at high masses above 800 GeV which will be accessible at $\sqrt{s}=3000$ GeV. Overall results show that the signal observation is possible at integrated luminosities below 1000 $fb^{-1}$ in type 3, and 10 $ab^{-1}$ in type 4 in the mass ranges under study within the limits of the analysis performed in this work. A more detailed study with possibly full detector simulation and more realistic collider environment is needed for the final conclusions.

		
\end{document}